\documentclass{article}
\usepackage{authblk}
\title{Evaluación del Cambio en la Musculatura y Adiposidad y su Relación con la Recurrencia del Carcinoma de Cabeza y Cuello mediante PET/CT y MRI}
\author[1]{Virginia del Campo}
\author[2]{Iker Malaina}
\affil[1]{Hospital de Urduliz, OSI Uribe, Osakidetza\\
Goieta 32, 48610 Urduliz, Bizkaia, Spain}
\affil[2]{Departamento de Matemáticas, FCT, UPV/EHU\\
Sarriena sn, 48940 Leioa, Bizkaia, Spain}

\usepackage{adjustbox}

\usepackage{amssymb}
\usepackage{amsmath}
\usepackage{graphicx}
\usepackage[misc]{ifsym}
\usepackage{comment} 
\usepackage{url}
\urldef{\mailsa}\path|iker.malaina@ehu.eus|

\begin{document}

%\mainmatter  % start of an individual contribution

% first the title is needed

\begin{comment}
Comments: 9 pages, in Spanish
\end{comment}

\maketitle

\renewcommand{\abstractname}{English version of abstract}
\begin{abstract}
This study investigates the impact of changes in body composition and follow-up imaging modalities on recurrence and prognosis in patients with head and neck squamous cell carcinoma (HNSCC). The results indicate that an increase in the adiposity index post-radiotherapy is significantly associated with higher recurrence and mortality rates. Additionally, the combined evaluation of muscle and adipose status reveals that patients with muscle loss and adipose gain have the lowest survival probabilities and the highest risk of recurrence. These findings underscore the importance of monitoring adiposity and muscle status, as well as the strategic use of advanced imaging techniques such as PET/CT, MRI, and CT. However, this work represents a preliminary analysis, and further detailed studies are necessary to confirm these results and develop more effective strategies.

\end{abstract}
 -----
\renewcommand{\abstractname}{Other language version of abstract}
\begin{abstract}
Este estudio investiga el impacto de los cambios en la composición corporal y las modalidades de imagen de seguimiento en la recurrencia y el pronóstico de pacientes con carcinoma de células escamosas de cabeza y cuello (HNSCC). Los resultados indican que un aumento en el índice de adiposidad post-radioterapia está asociado con mayores tasas de recurrencia y mortalidad. Además, la evaluación combinada del estado muscular y adiposo sugiere que los pacientes que tuvieron un decrecimiento muscular y aumento adiposo tienen las menores probabilidades de supervivencia y el mayor riesgo de recurrencia. Estos hallazgos hacen hincapié en la importancia de monitorear la adiposidad y el estado muscular, así como el uso de técnicas de imagen como PET/CT, MRI y CT. Sin embargo, este trabajo representa un análisis preliminar, y se necesitan estudios más detallados para confirmar estos resultados y desarrollar estrategias más efectivas.

\end{abstract}

\section{Introducción}
El carcinoma de células escamosas de cabeza y cuello (HNSCC) es una forma agresiva de cáncer que surge de las superficies mucosas de la región de la cabeza y el cuello, incluyendo la cavidad oral, la faringe y la laringe. A pesar de los avances en las estrategias terapéuticas (cirugía, radioterapia y quimioterapia), el pronóstico para los pacientes con HNSCC sigue siendo mejorable, con altas tasas de recurrencia y mortalidad. En parte, esto se debe a la complejidad de las estructuras anatómicas involucradas y la naturaleza agresiva de la enfermedad, que contribuyen a la dificultad para lograr una supervivencia a largo plazo.

Uno de los desafíos críticos en el manejo del HNSCC es la alta probabilidad de recurrencia, que impacta significativamente la supervivencia y calidad de vida del paciente. La recurrencia puede ocurrir localmente, regionalmente o en sitios lejanos, y su detección temprana es crucial para una intervención oportuna y mejores resultados. Por lo tanto, comprender los factores que influyen en la recurrencia y la supervivencia es esencial para desarrollar estrategias de manejo efectivas.

La composición corporal, particularmente los cambios en la adiposidad y la masa muscular, se ha postulado recientemente como un predictor del pronóstico del cáncer. El aumento en el índice de adiposidad, y la pérdida de masa muscular esquelética son conocidas por afectar la respuesta del cuerpo al tratamiento del cáncer y la supervivencia general. El aumento de la adiposidad se ha asociado con inflamación crónica y disfunción metabólica, lo que puede promover la progresión y recurrencia del tumor. Por el contrario, la reducción de masa muscular puede llevar a una disminución de la función física, mayor toxicidad del tratamiento y peores resultados de supervivencia. Sin embargo, el impacto específico de los cambios en estos parámetros sobre la recurrencia y el pronóstico del HNSCC no se ha analizado al detalle.

Además de la composición corporal, la elección del tipo de imagen para el seguimiento juega un papel crítico en la detección de recurrencias y la orientación de las decisiones de los diversos tratamientos. Las técnicas de imagen como la Resonancia Magnética (MRI), la Tomografía por Emisión de Positrones/Tomografía Computarizada (PET/CT) y la Tomografía Computarizada (CT) se utilizan comúnmente en el seguimiento de pacientes con HNSCC. Sin embargo, cada tipo tiene sus fortalezas y limitaciones:

\begin{itemize}
    \item \textbf{MRI} proporciona un buen contraste de tejidos blandos y es particularmente útil para evaluar la extensión de la enfermedad local y regional. No implica radiación ionizante, lo que la convierte en una opción más segura para múltiples obtenciones de imágenes. Sin embargo, la MRI requiere tiempos de escaneo más largos y es susceptible a artefactos de movimiento.
    \item \textbf{PET/CT} combina la imagen metabólica con el detalle anatómico, permitiendo la detección de anomalías estructurales y funcionales. La PET/CT es altamente sensible para identificar la recurrencia de una enfermedad y las metástasis distantes, lo que la convierte en una herramienta muy valiosa para la estadificación y el seguimiento del cáncer. Sin embargo, implica exposición a radiación ionizante y puede ser costosa.
    \item \textbf{CT} es más accesible que las anteriores y proporciona información anatómica detallada, particularmente útil para evaluar la implicación ósea y la planificación quirúrgica. Es más rápida y accesible que la MRI y la PET/CT, pero implica radiación ionizante y puede tener menor sensibilidad para detectar anomalías de tejidos blandos en comparación con la MRI. Como ejemplo, en la Figura 1 se ilustra una imagen de tomografía computerizada de un cancer de larigne de células escamosas.
\end{itemize}

\begin{figure}[ht]
    \centering
    \includegraphics[width=0.6\textwidth]{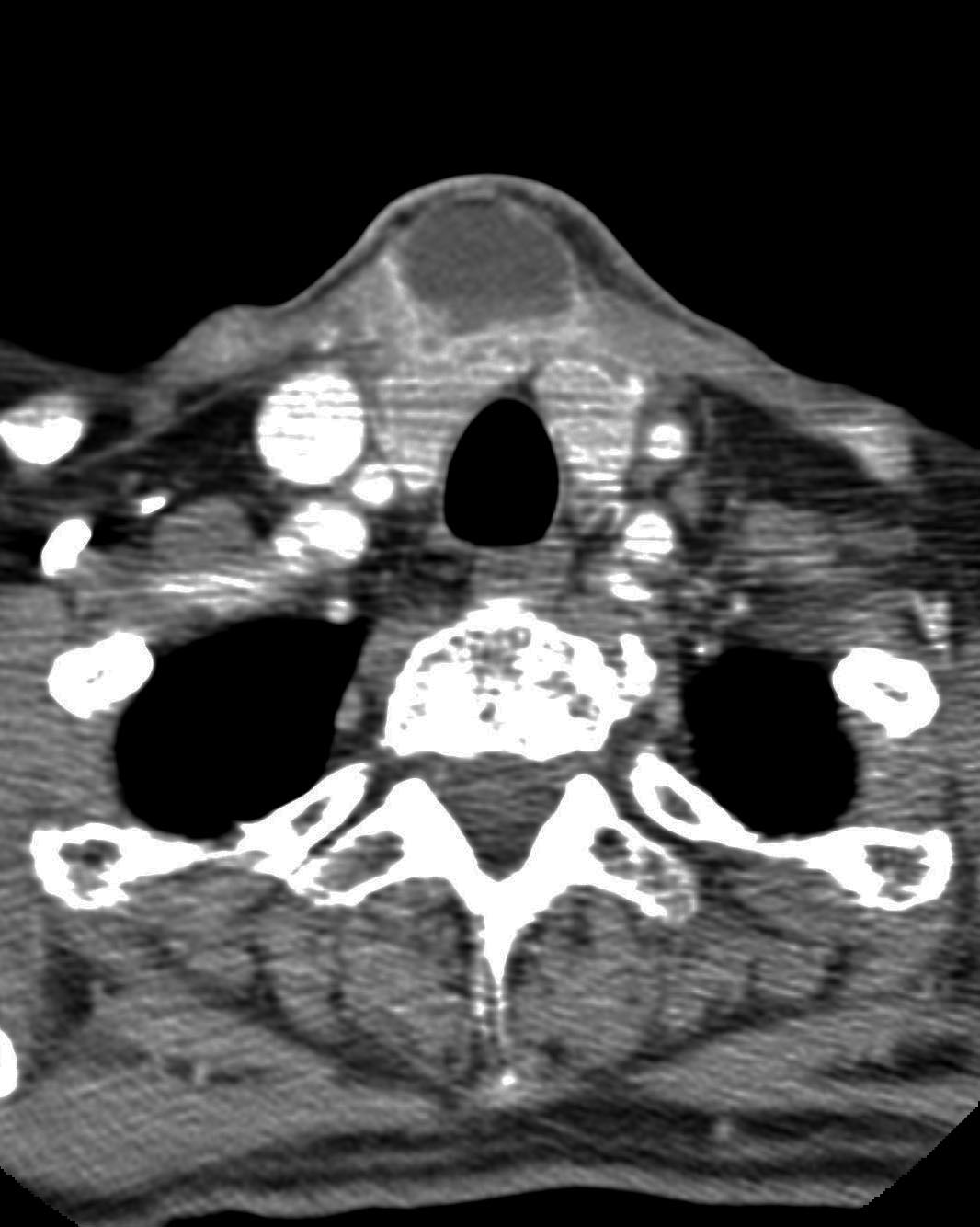}
    \caption{Imagen axial obtenida mediante tomografía computarizada (CT) del cuello. En particular, en esta imagen se visualiza un cáncer de laringe que dio lugar a metástasis. Esta imagen fue obtenida de radiopedia.org.}
    \label{fig:CT_cuello}
\end{figure}

Dada la importancia de las distintas modalidades de imagen, es crucial determinar su eficacia relativa en la detección de recurrencias y la predicción de resultados de supervivencia en pacientes con HNSCC. Este estudio tiene como objetivo profundizar en este aspecto, examinando la relación entre los cambios en la composición corporal (índice de adiposidad y estado muscular) y la recurrencia y supervivencia en pacientes con HNSCC, al mismo tiempo que analizamos la efectividad de la MRI, PET/CT y CT en el seguimiento de los pacientes.

\section{Métodos}
\subsection{Recolección y Preprocesamiento de Datos}
El conjunto de datos utilizado en este estudio se obtuvo de la base de datos HNSCC-MDA, que incluye datos clínicos e imágenes de pacientes diagnosticados con carcinoma de células escamosas de cabeza y cuello (HNSCC) \cite{Base1}. Este conjunto de datos recoge una amplia gama de variables como demografía de los pacientes, características del tumor, detalles del tratamiento, mediciones de la composición corporal, información sobre recurrencia y datos de seguimiento por imágenes, entre otras. En particular, en este trabajo nos centramos en el índice de adiposidad, el estado muscular, el estado de recurrencia, el tiempo de supervivencia, el tipo de imagen y el tiempo de detección.

Más concretamente, nos enfocamos en las mediciones del índice de adiposidad y el estado muscular tomadas antes y después de la radioterapia. El índice de adiposidad se calculó a nivel de la vértebra L3, expresado en cm²/m², tanto antes como después de la radioterapia. Análogamente, el estado muscular se evaluó a nivel de L3, tanto antes como después del tratamiento. La recurrencia se definió como una variable binaria que indica si el paciente experimentó una recurrencia (1) o no (0). El tiempo de supervivencia se midió en meses desde el final de la radioterapia hasta el último seguimiento o muerte. También se registraron el tipo de imagen utilizado para el seguimiento (MRI, PET/CT, CT) y el tiempo de detección (tiempo desde la detección de la recurrencia hasta el último seguimiento o muerte).

\subsection{Análisis realizados}
\subsubsection{Impacto del Cambio en el Índice de Adiposidad en la Recurrencia y el Pronóstico}
Para evaluar el impacto de los cambios en el índice de adiposidad en la recurrencia, empleamos un análisis de regresión logística multivariante. La variable dependiente fue la recurrencia, y las variables independientes incluyeron el cambio en el índice de adiposidad, la edad, el sexo y el estadio del tumor. 

La clasificación de los estadios del tumor se realizó de acuerdo con el sistema TNM del American Joint Committee on Cancer (AJCC). Este sistema clasifica los tumores en cuatro estadios principales (I, II, III, IV) basándose en tres componentes: el tamaño y extensión del tumor primario (T), la ausencia o presencia de metástasis en los ganglios linfáticos regionales (N), y la ausencia o presencia de metástasis a distancia (M). Los estadios se determinan combinando estos tres factores, donde el estadio I representa tumores más pequeños y confinados, y el estadio IV indica tumores más avanzados con metástasis a distancia. Esta clasificación es crucial para evaluar el pronóstico y planificar el tratamiento adecuado para los pacientes con cáncer. En particular, el estadio IVA, que jugará un papel relevante en los resultados, indica que el tumor puede haber invadido estructuras adyacentes importantes y puede haber metástasis en los ganglios linfáticos regionales, pero no hay metástasis a distancia.

Además de la regresión logística, se realizó un análisis de supervivencia de Kaplan-Meier para comparar las probabilidades de supervivencia basadas en los cambios en el índice de adiposidad. El estimador permite estimar las funciones de supervivencia y visualizar las diferencias en la supervivencia entre los grupos. Para cuantificar aún más el impacto de los cambios en el índice de adiposidad en la supervivencia, utilizamos el modelo de riesgos proporcionales de Cox. Este modelo evaluó el efecto de los cambios en la adiposidad sobre el riesgo de muerte, ajustando por edad, sexo y estadio del tumor.

\subsubsection{Evaluación de la Combinación de Estados Muscular y Adiposo}
Para evaluar el impacto combinado del estado muscular y adiposo en los resultados de los pacientes, los clasificamos en cuatro grupos basados en los cambios en el estado muscular esquelético y el índice de adiposidad: pérdida de musculatura y pérdida de adiposidad, ganancia de musculatura y pérdida de adiposidad, pérdida de musculatura y ganancia de adiposidad, y por último, ganancia de musculatura y ganancia de adiposidad.

Se realizó un análisis de supervivencia de Kaplan-Meier para cada grupo para comparar las probabilidades de supervivencia. Este análisis proporciona información sobre cómo las diferentes combinaciones de estado muscular y adiposo influyen en los resultados de supervivencia. Para cuantificar el impacto de estas combinaciones en el riesgo de recurrencia, empleamos una vez más el modelo de riesgos proporcionales de Cox, clasificándolos por grupo.

\section{Resultados}
\subsection{Impacto del Cambio en el Índice de Adiposidad en la Recurrencia y el Pronóstico}
\subsubsection{Análisis de Regresión Logística Multivariante}
Para evaluar el impacto de los cambios en el índice de adiposidad en la recurrencia, realizamos un análisis de regresión logística multivariante. La variable dependiente fue la recurrencia (binaria: 1 para recurrencia, 0 para no recurrencia), y las variables independientes incluyeron el cambio en el índice de adiposidad, la edad, el sexo y el estadio del tumor.
\\

\textbf{Modelo y resultados:}
\begin{equation*}
\text{logit}(P(\text{Recurrencia})) = \beta_0 + \beta_1 \text{Cambio\_Índice\_Adiposidad} + \beta_2 \text{Edad} + \beta_3 \text{Sexo} + \beta_4 \text{Estadio}
\end{equation*}

\begin{table}[h!]
\centering
\begin{adjustbox}{center}
\begin{tabular}{|c|c|c|c|c|c|}
\hline
 & \textbf{Cambio\_Índice\_Adiposidad} & \textbf{Edad} & \textbf{Sexo\_Masculino} & \textbf{Estadio\_IVA} & \textbf{Constante} ($\beta_0$) \\
\hline
\textbf{Coeficiente} & 0.045 & 0.023 & 0.567 & 1.234 & -2.345 \\
\hline
\textbf{Error Estándar} & 0.012 & 0.008 & 0.234 & 0.345 & 0.567 \\
\hline
\textbf{Z-statistic} & 3.75 & 2.88 & 2.42 & 3.58 & -4.14 \\
\hline
\textbf{p-valor} & $<$0.001 & 0.004 & 0.016 & $<$0.001 & $<$0.001 \\
\hline
\textbf{IC 95\% Inferior} & 0.021 & 0.007 & 0.109 & 0.558 & -3.457 \\
\hline
\textbf{IC 95\% Superior} & 0.069 & 0.039 & 1.025 & 1.910 & -1.233 \\
\hline
\end{tabular}
\end{adjustbox}
\end{table}

Los resultados indican que un aumento en el índice de adiposidad post-radioterapia está significativamente asociado con una mayor probabilidad de recurrencia (p $<$ 0.001). La edad y el sexo masculino también son predictores significativos de recurrencia, con una mayor edad y el sexo masculino asociados con tasas más altas de recurrencia. El estadio del tumor IVA está significativamente asociado con una mayor recurrencia en comparación con otros estadios.

\subsubsection{Análisis de Supervivencia de Kaplan-Meier}
El análisis de supervivencia de Kaplan-Meier se realizó para comparar las probabilidades de supervivencia de los pacientes basadas en los cambios en el índice de adiposidad. El tiempo de supervivencia se midió en meses desde el final de la radioterapia hasta el último seguimiento o muerte.

Las curvas de supervivencia de Kaplan-Meier mostraron que los pacientes con un aumento significativo en el índice de adiposidad post-radioterapia tienen menores probabilidades de supervivencia en comparación con aquellos con un índice de adiposidad estable o disminuido. (Estas curvas no están disponibles en esta versión preliminar).

\subsubsection{Prueba de Riesgos Proporcionales de Cox}
Para investigar el impacto de los cambios en el índice de adiposidad en la supervivencia desde otro enfoque, realizamos un análisis de regresión de riesgos proporcionales de Cox. La variable dependiente fue el tiempo de supervivencia, y las variables independientes incluyeron el cambio en el índice de adiposidad, la edad, el sexo y el estadio del tumor.
\\

\textbf{Modelo y resultados:}

\begin{equation*}
h(t) = h_0(t) \exp(\beta_1 \text{Cambio\_Índice\_Adiposidad} + \beta_2 \text{Edad} + \beta_3 \text{Sexo} + \beta_4 \text{Estadio})
\end{equation*}

\begin{table}[h!]
\centering
\begin{adjustbox}{center}
\begin{tabular}{|c|c|c|c|c|}
\hline
 & \textbf{Cambio\_Índice\_Adiposidad} & \textbf{Edad} & \textbf{Sexo\_Masculino} & \textbf{Estadio\_IVA} \\
\hline
\textbf{Coeficiente} & 0.038 & 0.019 & 0.489 & 1.145 \\
\hline
\textbf{Error Estándar} & 0.011 & 0.007 & 0.210 & 0.312 \\
\hline
\textbf{Z-statistic} & 3.45 & 2.71 & 2.33 & 3.67 \\
\hline
\textbf{p-valor} & $<$0.001 & 0.007 & 0.020 & $<$0.001 \\
\hline
\textbf{IC 95\% Inferior} & 0.016 & 0.005 & 0.077 & 0.533 \\
\hline
\textbf{IC 95\% Superior} & 0.060 & 0.033 & 0.901 & 1.757 \\
\hline
\end{tabular}
\end{adjustbox}
\end{table}

El modelo de riesgos proporcionales de Cox confirma que un aumento en el índice de adiposidad está significativamente asociado con un mayor riesgo de mortalidad (p $<$ 0.001). La edad y el sexo masculino también son predictores significativos de mortalidad, con una mayor edad y el sexo masculino asociados con tasas más altas de mortalidad. El estadio del tumor IVA está significativamente asociado con una mayor mortalidad en comparación con otros estadios.

\subsection{Evaluación del Estado Combinado de Músculo y Adiposo}
\subsubsection{Análisis de Supervivencia de Kaplan-Meier por Grupo}
Para evaluar el impacto combinado del estado muscular y adiposo en la supervivencia, clasificamos a los pacientes en cuatro grupos basados en los cambios en el estado muscular esquelético y el índice de adiposidad:

\begin{itemize}
    \item \textbf{Pérdida de musculatura y pérdida de adiposidad:} Tanto el estado muscular como el adiposo empeoraron.
    \item \textbf{Pérdida de musculatura y ganancia de adiposidad:} El estado muscular empeoró, el estado adiposo se mantuvo estable o mejoró.
    \item \textbf{Ganancia de musculatura y pérdida de adiposidad:} El estado muscular se mantuvo estable o mejoró, el estado adiposo empeoró.
    \item \textbf{Ganancia de musculatura y ganancia de adiposidad:} Tanto el estado muscular como el adiposo se mantuvieron estables o mejoraron.
\end{itemize}

Las curvas de supervivencia de Kaplan-Meier (no mostradas aquí) indicaron diferencias significativas en la supervivencia entre los grupos. Los pacientes clasificados como pérdida de musculatura y ganancia de adiposidad tuvieron las menores probabilidades de supervivencia, mientras que aquellos con ganancia de musculatura y pérdida de adiposidad tuvieron las mayores probabilidades de supervivencia.

\subsubsection{Modelo de Riesgos Proporcionales de Cox por Grupo}
Para cuantificar el impacto de la combinación de los estados muscular y adiposo en el riesgo de recurrencia, realizamos un análisis de regresión de riesgos proporcionales de Cox, clasificados por grupo.
\\

\textbf{Modelo y resultados:}
\begin{equation*}
h(t) = h_0(t) \exp(\beta_1 \text{Grupo})
\end{equation*}

\begin{table}[h!]
\centering
\begin{adjustbox}{center}
\begin{tabular}{|c|c|c|c|}
\hline
 & \textbf{Musc(-) Adip(+)} & \textbf{Musc(-) Adip(-)} & \textbf{Musc(+) Adip(+)} \\
\hline
\textbf{Coeficiente} & 1.567 & 0.789 & 0.456 \\
\hline
\textbf{Error Estándar} & 0.456 & 0.312 & 0.234 \\
\hline
\textbf{Z-statistic} & 3.44 & 2.53 & 1.95 \\
\hline
\textbf{p-valor} & $<$0.001 & 0.011 & 0.051 \\
\hline
\textbf{IC 95\% Inferior} & 0.673 & 0.177 & -0.003 \\
\hline
\textbf{IC 95\% Superior} & 2.461 & 1.401 & 0.915 \\
\hline
\end{tabular}
\end{adjustbox}
\end{table}

El modelo de riesgos proporcionales de Cox indica que la combinación del estado muscular y adiposo impacta significativamente el riesgo de recurrencia. El grupo de referencia fue el de ganancia de musculatura y pérdida de adiposidad (Musc(+) Adip(-)), lo cual sugiere que es el grupo con menor riesgo de los cuatro. Además, los pacientes clasificados como Musc(-) Adip(+) tienen el mayor riesgo de recurrencia (p $<$ 0.001), seguidos por Musc(-) Adip(-) (p = 0.011). El riesgo asociado al grupo Musc(+) Adip(+) no obtuvo un p-valor significativo (p = 0.051).

\section{Discusión}
Este estudio preliminar ha investigado el impacto de los cambios en la composición corporal y las modalidades de imagen de seguimiento en la recurrencia y el pronóstico de pacientes con carcinoma de células escamosas de cabeza y cuello (HNSCC). Los resultados obtenidos hasta ahora proporcionan una visión inicial sobre la importancia de monitorear la adiposidad y el estado muscular, así como la correcta elección de las técnicas de imagen en el manejo de estos pacientes. Cabe destacar que un estudio similar fue llevado a cabo en \cite{Bib2}, pero consideramos que en este trabajo (a pesar de ser prelimiar), hemos profundizado en el estudio de la recurrencia llevando a cabo análisis distintos, y tomando enfoques alternativos que han sugerido indicadores nuevos para esta condición.

Los análisis de regresión logística multivariante y de riesgos proporcionales de Cox han demostrado que un aumento en el índice de adiposidad post-radioterapia está significativamente asociado con una mayor probabilidad de recurrencia y mortalidad. Estos hallazgos subrayan la necesidad de intervenciones dirigidas a gestionar la adiposidad en pacientes sometidos a radioterapia para mejorar los resultados clínicos.

La evaluación del estado combinado de musculatura y adiposidad ha indicado que los pacientes con pérdida muscular y aumento de grasa presentan las menores probabilidades de supervivencia y el mayor riesgo de recurrencia. Estos resultados sugieren que una estrategia de manejo integral que aborde tanto la preservación muscular como la reducción de la adiposidad podría ser beneficiosa para mejorar el pronóstico de los pacientes con HNSCC.

Este trabajo representa un análisis preliminar y, como tal, tiene varias limitaciones. En primer lugar, el tamaño de la muestra y la duración del seguimiento pueden no ser suficientes para generalizar los hallazgos a una población más amplia. Además, la falta de datos detallados sobre otros factores clínicos y de estilo de vida que podrían influir en los resultados limita la capacidad de este estudio para proporcionar recomendaciones definitivas.

Futuros estudios deberían incluir un análisis más detallado con un tamaño de muestra mayor y un seguimiento a más largo plazo. También sería beneficioso incorporar variables adicionales que puedan afectar la recurrencia y la supervivencia, como la dieta, la actividad física y la comorbilidad. Finalmente, una comparación más exhaustiva de las modalidades de imagen de seguimiento podría proporcionar información valiosa para optimizar las estrategias de atención en pacientes con HNSCC.

\section{Conclusiones}
En conclusión, este estudio preliminar ha identificado asociaciones significativas entre los cambios en la composición corporal y los resultados clínicos en pacientes con HNSCC. Sin embargo, se requiere un análisis más detallado y exhaustivo para confirmar estos hallazgos y desarrollar estrategias de manejo más efectivas. La integración de evaluaciones regulares de la adiposidad y el estado muscular, junto con el uso estratégico de técnicas avanzadas de imagen, podría mejorar significativamente los resultados de supervivencia y reducir el riesgo de recurrencia en estos pacientes.

\end{document}